\documentclass[draft]{agujournal2018}
\usepackage{apacite}
\usepackage{url} 
\usepackage{enumitem}
\usepackage{amsmath}
\usepackage{overpic}
\usepackage{pict2e}
\usepackage{bm}

\draftfalse

\journalname{Journal of Geophysical Research}

\journalname{JGR-Atmospheres}

\begin{document}

%
%


\title{
Tracking the Gust Fronts of Convective Cold Pools
}

%
%




\authors{Marielle B. Fournier and Jan O. Haerter}


\affiliation{1}{Niels Bohr Institute, University of Copenhagen, Blegdamsvej 17, 2100 Copenhagen, Denmark}




\correspondingauthor{Jan O. Haerter}{haerter@nbi.ku.dk}




\begin{keypoints}
\item A simple tracking algorithm for simulated cold pool gust fronts is implemented;
\item large eddy simulations are used to validate the algorithm;
\item a setup-independent relation between precipitation intensity and cold pool momentum is found.
\end{keypoints}

%
%


\begin{abstract}
It is increasingly acknowledged that cold pools can influence the initiation of new convective cells. Yet, the full complexity of convective organization through cold pool interaction is poorly understood. This lack of understanding may partially be due to the intricacy of the dynamical pattern formed by precipitation cells and their cold pools. Additionally, how exactly cold pools interact is insufficiently known. To better understand this dynamics, we develop a tracking algorithm for cold pool gust fronts. Rather than tracking thermodynamic anomalies, which do not generally coincide with the gust front boundaries, our approach tracks the dynamical cold pool outflow. Our algorithm first determines the locus of the precipitation event. Second, relative to this origin and for each azimuthal bin, the steepest gradient in the near-surface horizontal radial velocity $v_r$ is employed to determine the respective locus of the cold pool gust front edge. 
Steepest $v_r$-gradients imply largest updraft velocities, hence strongest dynamical triggering. 
Results are compared to a previous algorithm based on the steepest gradient in temperature --- highlighting the benefit of the method described here in determining dynamically active gust front regions. 
Applying the method to a range of numerical experiments, the algorithm successfully tracks an ensemble of cold pools. 
A linear relation emerges between the peak rain intensity of a given event and maximal $v_r$ for its associated cold pool gust front --- a relation found to be nearly independent of the specific sensitivity experiment.
\end{abstract}

\section{Introduction}
\noindent
As rain falls towards the ground, a fraction of it is evaporated into the unsaturated sub-cloud layer \citep{srivastava1987model,li2001analytical,seifert2008parameterization, lolli2017rain}. 
The resulting evaporatively cooled and therefore relatively dense, volume is often referred to as a {\it cold pool}.
The gravitational force due to the density increase, along with downdrafts, causes the air to descend and spread out horizontally along the surface. 
This outward propagation has been shown to resemble that of density currents \citep{charba1974application}. 
As the cold air spreads, the surrounding warmer air is forced upwards leading to strong, positive vertical velocities at the cold pool gust fronts, which can trigger convection and the formation of clouds along these fronts \citep{purdom1976some}.

\noindent
Both observational and numerical studies have examined this triggering effect of cold pools \citep{rotunno1988theory,khairoutdinov2006high,li2014simulated,jeevanjee2015effective,feng2015mechanisms}. 
They find that the lifting of environmental air occurs either due to the interaction of the cold pool gust front with the environmental winds, typical of squall lines and long-lived multicellular thunderstorms \citep{rotunno1988theory,weisman1988structure,torri2015mechanisms,he2018initiation}, or due to the collision of gust fronts \citep{feng2015mechanisms,droegemeier1985three,kingsmill1995convection,lima2008convective}.

\noindent
Apart from this dynamical triggering, the formation of cold pools also alters the thermodynamic properties near the surface \citep{terai2013aircraft,de2017cold,zuidema2017survey}. 
The cooler air increases the static stability leading to local suppression of convection inside the cold pools. Together with the enhanced wind speeds associated with the outflow, the cooler air influences the energy exchange between the surface and the overlaying air by altering the surface fluxes \citep{schlemmer2016modifications,gentine2016role,grant2018cold}. 
Cold pools are sometimes surrounded by areas of positive moisture perturbations \citep{tompkins2001organization,schlemmer2014formation}, which can aid in the triggering of new convection at the cold pool gust fronts by locally lowering the convective inhibition (CIN) and increasing the buoyancy.

\noindent
The convective triggering associated with cold pools affects the spatial and temporal distribution of the resulting cloud field, organizing it into clusters surrounding the cold pools, which in turn affects the radiative budget of the atmosphere. 
It has been shown that cold pools are important for the transition from shallow to deep convection over land \citep{khairoutdinov2006high,boing2012influence,schlemmer2014formation}, where the formation of deeper clouds and the peak in rain intensity does not develop until late in the afternoon, despite the presence of large values of convective available potential energy (CAPE) \citep{nesbitt2003diurnal}. 
Since convection is not fully resolved in large scale climate models, they heavily rely on quantities such as CAPE in order to simulate convection. 
A symptom of this simplification is that deep convection develops too early in the diurnal cycle \citep{betts2002evaluation}.

\noindent
Addressing this shortcoming, \citet{rio2009shifting} together with \citet{grandpeix2010density} introduced a cold pool model into the parameterization of convection in a single column version of a general circulation model (GCM), and could partially remedy the shortcoming by simulating a more realistic diurnal cycle of convective precipitation. \citet{khairoutdinov2006high} showed, by switching off evaporation of precipitation (removing cold pools) in a cloud-resolving model, that cold pools were needed in order to generate large enough thermals to support the growth of deep clouds.

\noindent
Doubtlessly, despite these advances, there is further need for better understanding of the dynamics and evolution of cold pools and how they act to organize convection. 
The spatial distribution and subsequent convective triggering of cold pools can be examined by tracking how cold pools evolve throughout the day. 
This motivates the use of a tracking algorithm that can identify and follow cold pools throughout their entire lifetime. 
Given that the triggering of new convective events predominantly occurs at or near the cold pool gust fronts, these loci of strong horizontal convergence must be determined by such a tracking algorithm. 

Other tracking methods have been developed in recent years: \citet{schlemmer2014formation}, \citet{torri2015mechanisms} as well as \citet{gentine2016role} used thresholds on temperature anomalies to identify cold pools and subsequent spatially connected regions. 
Similarly, using proxies for buoyancy
, such as density potential temperature, $\theta_{\rho}$, 
\citet{tompkins2001organization} and \citet{feng2015mechanisms} determined the spatial extent of the cold pools either subjectively \citep{tompkins2001organization} or by using an image processing technique \citep{feng2015mechanisms}. 
Common to these studies is that the thresholds are determined subjectively by visual inspection of the corresponding fields. 
A recent study by \citet{drager2017characterizing} sought to eliminate such subjective thresholds by using gradients in $\theta_{\rho}$ rather than absolute values. 
Cold pools were thereby identified as closed boundaries defined by the zero contours of the second radial derivative of $\theta_{\rho}$.

\begin{figure}[htb]
    \centering
   \includegraphics[width=.9\textwidth]{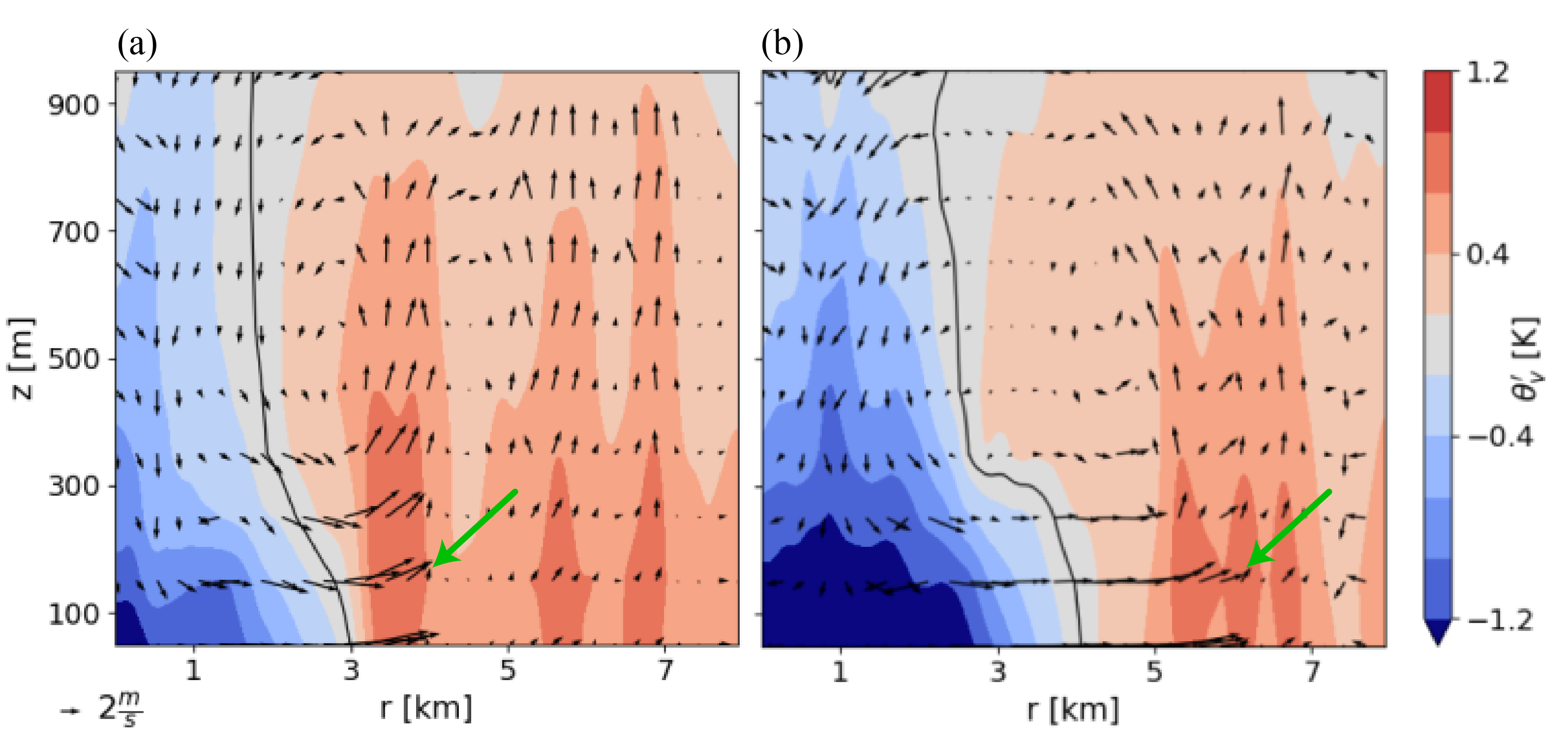}
    \caption{{\bf Cross section through a simulated cold pool.} 
    The vertical and horizontal axes show height and radial position relative to the cold pool center, respectively; 
    (a) and (b) correspond to times 15 and 30 minutes after cold pool identification. 
    Color shades indicate virtual potential temperature anomaly $\theta_v'$ ({\it see} color bar), solid black lines denote $\theta_v'$=$0$ and arrows the $r$---$z$ wind field. 
    The green arrows highlight the loci of maximum updrafts at $z=200\;m$.
    A reference wind vector is shown in the lower left corner of panel (a).
    Note that the radial position of maximum updrafts lies further away from the center than that of the zero of $\theta_v'$ in both panels, and that it advances more rapidly.  Data: simulation p2K (see Sec.~\ref{sec:methods}).
    }
    \label{fig:cross}
\end{figure}

\noindent
These tracking methods share the use of a thermodynamic quantity as cold pool identifier.
Such an approach warrants the forcing, that is, potential energy, which could drive a cold pool initially at rest. 
However, especially during the final stages of the cold pool lifetime, cold pools expand due to the inertia gained much earlier and the air near the gust fronts may no longer be anomalously dense: 
during the continuous spreading of a cold pool, the gust front experiences turbulent mixing and entrainment of environmental air. 
As a result, surface energy fluxes will act to increase the overall buoyancy of the cold pool. 
A recent study by \citet{grant2018cold} indeed shows that this effect is most pronounced near the cold pool gust fronts and that cold pool dissipation therefore proceeds from the outer edge inward. 
These factors reduce the difference in temperature and density between the cold pool and the environment. 
Despite this gradual thermodynamic equilibration process, the air at the cold pool edges will still maintain its inertia or could even be additionally forced outwards by the dense air masses in the cold pool interior.

To be more specific, consider an example (Fig. \ref{fig:cross}). 
The updraft associated with the cold pool gust front has advanced further than the temperature anomaly --- a discrepancy explained by enhanced surface-to-atmosphere energy fluxes under the strong horizontal near-surface winds, as well as turbulent mixing within the cold pool gust front \citep{tompkins2001organization,schlemmer2014formation}. 
In short: if one were to identify the cold pool using density measures, one would generally not detect the locus of convergence, where triggering of new convection is expected.
Our current method remedies this issue.

The structure of this paper is as follows. 
We first describe our test data (Sec.~\ref{sec:methods}) and tracking algorithm (Sec. \ref{sec:cp_tracking_algorithm}), and then apply the tracking method based on the dynamical aspects of cold pools (Sec.~\ref{sec:results}).
We discuss the method in terms of algorithm performance, including a comparison with the thermodynamic-based tracking algorithm developed by \citet{drager2017characterizing} ({\it hereafter:} DH17) and the temporal evolution during a simulated diurnal cycle. 
Sec.~\ref{sec:conclusion} concludes and offers examples of where the method could usefully be applied.

\section{Materials and Methods}\label{sec:methods}
\noindent
To test the method, we use an idealized diurnal cycle simulation, which mimics mid-latitude summertime convection. 
This transiently varying simulation setup was chosen, as it generates cold pools of various spreading velocities and length scales, which vary over the course of the day \cite{haerter2017precipitation}.

\noindent
\subsection{Simulation Setup}

\noindent
The convective atmosphere was simulated using the University of California, Los Angeles (UCLA) Large Eddy Simulation (LES) model with sub-grid scale turbulence parameterized using the Smagorinsky model, a delta four-stream radiation scheme and a two-moment cloud microphysics scheme \citep{stevens2005evaluation}. 
Rain evaporation is implemented after \citet{seifert2008parameterization}.
As detailed in \citet{moseley2016intensification}, diurnally oscillating surface temperature ($T_s(t)$) boundary conditions are applied, with

\begin{eqnarray}
  T_s(t)=\overline{T_{s}}-T_{a} \cos{2\pi \,t/t_0}\;,
\end{eqnarray}

\noindent
with $\overline{T_{s}}$ the average surface temperature, $T_{a}=10\;K$ the surface temperature amplitude and $t_0$ the duration of the simulated model day.
All simulations were initialized with data from observed summertime mid-latitude conditions where convection had occurred in order to establish an initially unstable atmosphere. 
As in \citet{moseley2016intensification} we vary 
\begin{itemize}
\item $\overline{T_{s}}\in\{23,25,27\}\;^{\circ}C$, yielding simulations denoted as CTR, p2K and p4K, respectively, 
\item keeping $\overline{T_{s}}$ fixed to $23\;^{\circ}C$, varying  $t_0\in\{1,2\}\,day$ modifies the buoyant instability and the duration, over which cold pools can organize, respectively.
The "longer day" simulation with $t_0=2\,day$ is referred to as LD.
\end{itemize}

\noindent
Increasing $\overline{T_s}$, but not the atmospheric initial conditions, corresponds to greater convective instability due to the adjustment time required for the atmosphere to reach equilibrium.

\noindent
The model numerically integrates the anelastic equations of motion on a regular horizontal domain ({\it see} Tab. \ref{tab:summary} for domain sizes) with 200 m horizontal grid spacing and periodic boundary conditions. 
The model spans 75 stretchable vertical levels with spacings: 100 m below 1 km height, 200 m between 1 and 6 km and 400 m between 6 km and the domain top, located at 16.5 km. 
Additionally, the model uses a sponge layer above 12.3 km. 
The horizontal domain sizes (Tab. \ref{tab:summary}) were chosen in order to obtain sufficient statistics to distinguish the effect of the different surface boundary conditions.

The Coriolis force and the mean wind were set to zero with weak random initial perturbations added as noise to break complete translational symmetry. 
No large scale forcing was imposed, ensuring that the only driving force for convection was buoyancy and the forced lifting through cold pool interaction.
The model output time step was $\Delta t_{out}=5\;min$. 
At each output time step, $5\;min$-accumulated precipitation and instantaneous velocities were output at each model gridbox. 

\noindent
\section{Cold Pool Tracking Algorithm}\label{sec:cp_tracking_algorithm}
\noindent
\subsection{Dynamical approach}

\begin{figure}[htb]
    \centering
   \includegraphics[width=.6\textwidth]{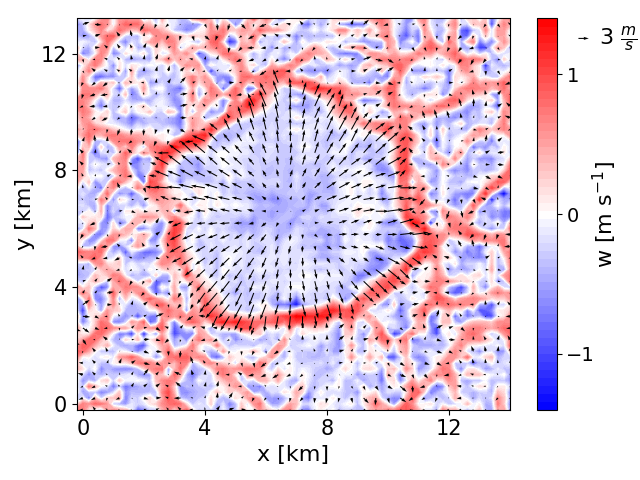}
    \caption{{\bf Typical wind field of a simulated cold pool.} Colors denote near-surface ($z = 100\;m$) vertical velocity while the near-surface horizontal wind field ($z=50m\;$) is depicted by arrows. The reference arrow size is found in the upper right corner. Data: simulation p2K (see Sec.~\ref{sec:methods}).}
    \label{fig:cpwind}
\end{figure}

\noindent
Cold pools are characterized by approximately circular expansion of the near-surface horizontal wind field, with winds directed radially outward from the precipitation event (Fig. \ref{fig:cpwind}):
horizontal velocity vectors point in radial directions and, due to the anelastic continuity equation,

\begin{equation}
    \frac{\rho_{0}}{r}\frac{\partial(rv_{r})}{\partial r} + \frac{\partial (w\rho_{0})}{\partial z} = 0\;,
    \label{eq:masscon}
\end{equation}

\noindent
with $\rho_0$ the mean state air density and $r$ the radial coordinate, the magnitude of radial velocities decreases in locations where vertical velocity is largest. 
In the Fig. \ref{fig:cpwind}, these locations are seen as a pronounced red ring of approximately $500$ $m$ thickness. 
This ring constitutes a demarcation line between the organized, strong, wind pattern within, and less organized, weaker wind outside of this line.
Additionally, note that the vertical wind in the interior of this ring is generally negative, corresponding to downdraft regions where further convection is generally inhibited.
Virtual potential temperature anomaly $\theta_v'$ is often used as a measure of buoyancy. 
Comparing again with Fig.~\ref{fig:cross}, where the line $\theta_v'=0$ lies one or several $km$ closer to the horizontal origin than the locus of maximum updraft, it is clear that $\theta_v'=0$ would often identify regions of suppressed, rather than favored convection. 

To detect locations of strong gust front vertical velocity, at each time step our algorithm hence locates points of sharpest decrease in radial velocity, hence, largest resulting vertical velocity. 
In the following, we denote these locations as \textit{edges}, and generally associate them with the actual cold pool gust fronts.

\noindent
\subsection{Algorithm}\label{sec:algo}

\noindent
In contrast to the cold pool gust fronts we want to detect, the precipitation events that cause these gust fronts are easily discernible, as they form sharp horizontal boundaries. 
Patches of surface precipitation can be identified as horizontally and temporally contiguous areas of nonzero surface precipitation. 
Cold pools systematically emanate from these patches, termed {\it parent rain events}, and each cold pool can therefore be uniquely associated with a specific parent rain event.
As will be shown, by using these parent rain events as the spatial and temporal reference for the corresponding cold pool detection, the gust fronts can be mapped out much more systematically. 

The advantage of considering precipitation and cold pools on the same footing, is twofold: 
(i) the relation between surface precipitation and the cold pool properties can be studied. 
(ii) the parent rain event serves as a natural origin, both in space an time, for the emergent cold pool. 

\noindent
The algorithm is split into three phases:
\begin{enumerate}
    \item Rain cell tracking,
    \item Computation of $v_{r}$ and $\partial v_{r}/\partial r$,
    \item Identification of cold pool edges based on the minimum in $\partial v_{r}/\partial r$.
\end{enumerate}

\noindent
{\bf 1. Rain cell tracking}.\\
\noindent
Rain cells are tracked using the iterative rain cell tracking (IRT) \citep{moseley2013probing}.
The IRT detects spatially contiguous rain events with surface rainfall rate above a set threshold on intensity, $I_{0}$, at each time step. 
Subsequently, {\it rain tracks}, temporally contiguous patches, are identified by considering grid box overlaps for the rain objects forwards and backwards in time.

\noindent
We find that rain events that cover very small surface areas do not produce detectable cold pools, and our method therefore makes use of a lower areal threshold of 50 contiguous grid cells.
This corresponds roughly to an area of 2 km$^{2}$ in this model setup. Smaller areal thresholds were tested, which resulted in more merging cold pools and not well defined cold pools in the radial velocity at the initial identification time.
For a gridbox to be considered rainy, we set the threshold $I_0=1\;mm\;h^{-1}$, as did DH17. \citet{zuidema2012trade} used $I_0=0\;mm\;h^{-1}$, 
while \citet{barnes1982subcloud} argue 
for $I_0=2\;mm\;h^{-1}$. 
We chose $I_0=1\;mm\;h^{-1}$, because the Rain in Cumulus over the Ocean (RICO) experiment \citep{rauber2007rain} showed that nearly all clouds with precipitation rates larger than $1$ $mm\;h^{-1}$ were associated with cold pool like outflow \citep{snodgrass2009precipitation}. 
Temporally, rain tracks with a total lifetime of less than 10 minutes were disregarded as visual inspection did not reveal well-defined cold pools in the near-surface surface $v_{r}$ field for these tracks.

\noindent
Rain tracks are detected for as long as the requirements on area and rain intensity are fulfilled. 
Usually, a time lag between the emergence of a rain event and the appearance of the associated cold pool in the surface $v_{r}$ field is observed. Therefore, the tracking of the cold pool begins in the succeeding time step ($\Delta t_{out}=5\;min$) of the initial identification of the rain event. 
Even after the parent rain track ceases, the corresponding cold pool is tracked as long as it is considered {\it dynamically active}. 
We find that there are many plausible options in terminating the track of a given cold pool. We here employ a threshold velocity on the mean radial velocity. Cold pools with inferior velocities are no longer considered active. As a pragmatic and transparent choice we used a threshold of $1$ $m\;s^{-1}$, which is significantly beyond the velocity fluctuations not caused by the cold pool gust fronts. For other studies lower or larger values could be more optimal depending on the research question.

\begin{figure}
       \includegraphics[width=.9\textwidth]{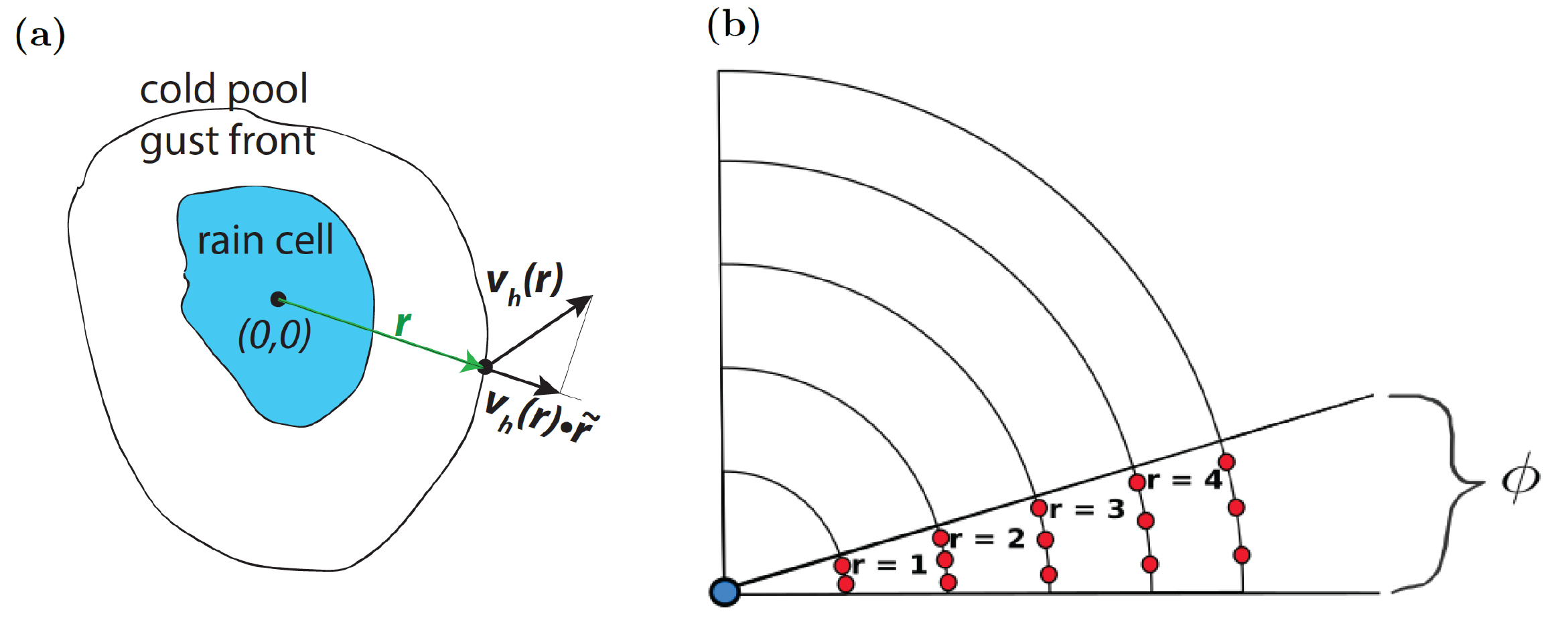}
    \caption{{\bf Definition of radial velocity, its computation and the cold pool partitioning used in detecting the edge.} 
    (a), Schematic showing a rain cell, its cold pool gust front, and the definition of position and velocity vectors, where
    $\bm{v}_h(\bm{r})$ is the horizontal velocity at the position $\bm{r}$.
    (b), Schematic illustrating one of the azimuthal intervals ($\phi=11.25^{\circ}$). 
    The blue filled circle denotes the identified COM and the red filled circles denote radial distance rounded to multiples of the grid spacing. 
    At each of these positions $\partial v_{r}(r)/\partial r$ is averaged over the given azimuthal interval.}
    \label{fig:azint}
\end{figure}

\noindent
{\bf 2. Computation of  $\bm{v_{r}}$ and $\bm{\partial v_{r}/\partial r}$}\\
\noindent
For each rain cell and each simulation output time step we define the origin $\bm{O}\equiv (0,0)$ as the precipitation intensity-weighted center of mass (COM). 
As the rain cell area generally changes from one time step to the next, its COM will also undergo corresponding displacements. 
At any position $\bm{r}$ relative to $\bm{O}$, we obtain the time dependent radial wind speed by computing the projection
\begin{equation}
v_{r}(\bm{r})\equiv \bm{v}_h(\bm{r})\cdot \bm{\hat{r}}\;,
\label{eq:v_r}
\end{equation}
for each horizontal point $\bm{r}\equiv (x,y)=r\cos\phi\;\hat{\bf x}+r\sin\phi\;\hat{\bf y}$ within the model domain, 
where $r\equiv \|\bm{r}\|$, $\bm{\hat{r}}\equiv \bm{r}r^{-1}$  is the unit vector in the direction of $\bm{r}$ (Fig.~\ref{fig:azint}a) and $\phi$ is the azimuth.
The velocity vector $\bm{v}_h(\bm{r})\equiv (v(r),u(r))$ is the horizontal velocity at the position $\bm{r}$, where $v$ and $u$ are the velocity components in the $x$- and $y$-directions.

It is now convenient to rotate the coordinate system so that the vector $\bm{r}$ is oriented along $\hat{\bm{x}}$.
This is accomplished by applying the $2\times 2$-rotation matrix $\mathcal{R}(-\phi)$ to all horizontal vectors. 
The radial unit vector $\hat{\bm{r}}$ is then mapped into $\hat{\bm{x}}$ and the derivative in the radial direction just becomes a scalar derivative.
The derivative is obtained for each $\bm{r}$ using central finite differences of fourth order accuracy
\begin{equation}
    \frac{\partial v_{r}(r)}{\partial r} \approx \frac{v_{r}(r-\Delta s) - 8v_{r}(r-2\Delta s) + 8v_{r}(r+2\Delta s) - v_{r}(r+\Delta s)}{12}\;,
\end{equation}
where $\Delta s$ is the model's grid spacing ($\Delta s=200\;m$ here, {\it see} Sec.~\ref{sec:methods}).
$v_{r}(r \pm \Delta s)$ and $v_{r}(r \pm 2\Delta s)$ are determined by bilinear interpolation between the
respective grid points.

As our current simulations do not employ external wind shear, the horizontal displacement of the origin from one time step to the next is small.
Yet, these small displacements often are important when calculating a radial quantity and searching for minima in derivatives --- we hence find it useful to consider them. 
Additionally, allowing for a time dependent center of mass makes the algorithm more versatile and applicable to simulations with different setups, such as boundary conditions with large scale wind shear, where the displacements of the COM are surely much larger.

\noindent
{\bf 3. Identification of cold pool edges based on the minimum in $\bm{\partial v_{r}/\partial r}$}\\
\noindent
After computing $v_{r}$ and $\partial v_{r}/\partial r$, our method now determines the positions of the cold pool edges. 
While cold pools are often approximated as perfectly circular objects \citep{grandpeix2010density,romps2016sizes}, circularity is a poor approximation as soon as they deform under collision, and the formation of new cold pools is triggered \citep{torri2019cold}. 
Thus, in order to correctly track all shapes, the area surrounding each cold pool is sub-divided into a unit circle centered at the cold pool COM, which is then split into 32 azimuthal intervals, termed {\it slices}, each spanning $\phi = 11.25^{\circ}$ (Fig. \ref{fig:azint}b). 
Different slice sizes were tested regarding the identification of a single cold pool and the one yielding edges encircling the maximum radial velocity of the cold pool was chosen.
The relatively small values of $\phi$ indicates considerable heterogeneity of the dynamical cold pool edge contour, which we mainly attribute to deformations from mostly circular shapes during the initial phase of spreading, to more Voronoi graph type structures upon collisions between cold pools \cite{haerter2019circling}.

Subsequently, the cold pool COM and the surrounding grid points are aligned in space by rounding the coordinate of its COM and $r$ to the nearest integer multiple of $\Delta s$ --- in effect constituting a radial binning. 
The binning ensures that $\partial v_{r}/\partial r$ can be smoothed in each of the resulting slices by averaging at each radius (Fig. \ref{fig:azint}b).
The edge is then identified in each slice by locating the radius with minimum $\partial v_{r}/\partial r$, termed $r^*(\phi_{j})$, where $\phi_{j}$ denotes the central azimuth of a given slice.
Note that one could equivalently detect zeros in $\partial^2 v_{r}/\partial r^2$ and require a negative third derivative.
Numerically, this option was however found more cumbersome.
The approach we employ here ensures that at least one edge point is detected in each slice and that a closed contour surrounding a given cold pool can always be mapped out.

\noindent
The location of the edge in one slice is constrained by the location of the edge in the previous slice by a {\it neighbor constraint}: 
this constraint limits the radial search range for $r^*(\phi_{j})$ to

\begin{equation}
    r^*(\phi_{j-1}) - dr \le r(\phi_{j}) \le r^*(\phi_{j-1}) + dr
    \label{eq:neighbourcon}
\end{equation}

\noindent
where $dr = 3\Delta s$. $dr$ is fitted to the model horizontal resolution used in this study and should be revised if used with another resolution.

\begin{figure}[htb]
    \centering
       \includegraphics[width=.6\textwidth]{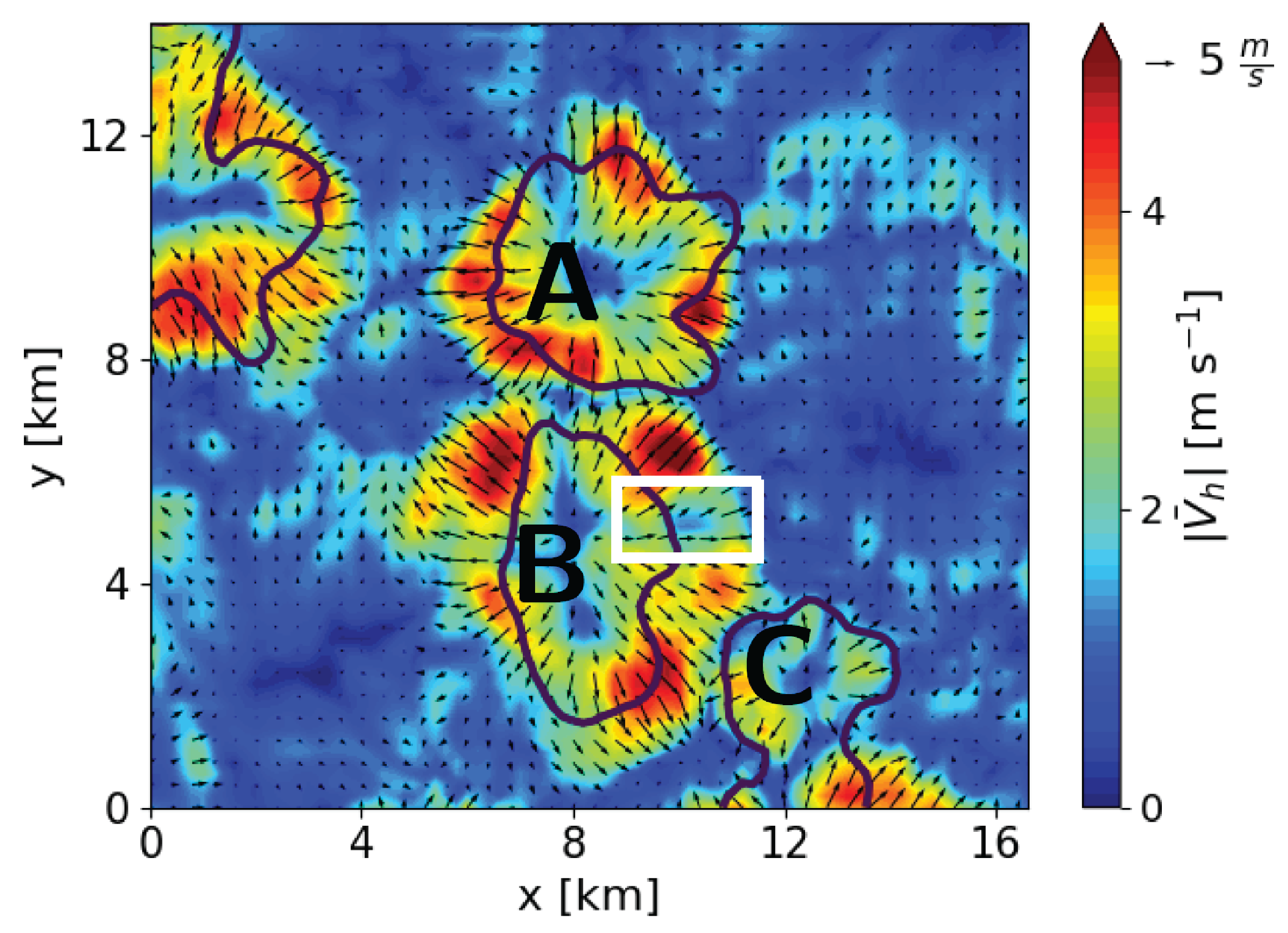}
    \caption{{\bf Horizontal wind field generated by a population of cold pools.} 
    Wind field magnitude and direction are shown by color shades and arrows, respectively. 
    Three cold pools are highlighted by A,B,C. 
    The dark purple lines denote the contour of the surface precipitation cutoff at $1\;mm\;h^{-1}$ at the same time step. 
    Note that the parent rain event of cold pool C is in the process of merging with that of the cold pool situated immediately below.
    The example illustrates the challenges that are sought overcome by the consistency checks mentioned in the text.
    The white box highlights an area where multiple maxima in $v_{r}$ are visible.
    Note the reference arrow, shown in the upper right outside the figure. Data: simulation p2K (see Sec.~\ref{sec:methods}).}
    \label{fig:cpconstraints}
\end{figure}

\noindent
{\bf Consistency checks.} Repeated checks of the following four constraints are performed after the initial identification of the edge in a slice.

\begin{enumerate}
    \item \textit{The edge must not be located at negative $v_{r}$:}
    \begin{equation}
        v_{r}(r^*) > 0
    \end{equation}
\end{enumerate}
We generally disregard contracting azimuthal slices (where $v_r<0$).
Examining cold pool A in Fig. \ref{fig:cpconstraints}, an edge identified where $v_{r}<0$ would imply that the edge is found in the interior of cold pool B. 
If this occurs, the algorithm disregards the identified edge and the radial search range determined by the neighbour-constraint (Eq. \ref{eq:neighbourcon}) is moved further towards the cold pool COM.
\begin{enumerate}[resume]
    \item \textit{$v_{r}$ at positions immediately surrounding the identified edge should not be larger than $v_{r}$ at the edge:}
    \begin{equation}
        v_{r}(r) \leq v_{r}(r^*) \hspace{0.2cm}for\hspace{0.2cm} r^* < r < r^* + dr
    \end{equation}
\end{enumerate}
where $dr$ was chosen to be the distance covered by three pixels in the respective azimuthal direction. $v_{r}$ does not increase monotonically from the cold pool COM to the edge (see white box in Fig. \ref{fig:cpconstraints}). Turbulent mixing and surface energy fluxes alter the cold pool as it expands, generally yielding multiple maxima in $v_{r}$ and minima in $\partial v_{r}/\partial r$. 
Limiting this constraint to only consider $r^* + dr$ was found useful, as larger limits led to identification of the edges of other cold pools situated in the vicinity e.g. cold pool C located close to cold pool B (Fig. \ref{fig:cpconstraints}). The constraint can potentially push the identified edge to larger radii for isolated cold pools due to velocity fluctuations in the immediate surroundings but will not be a problem between colliding cold pools. In the latter case, the immediate surroundings of the one cold pool will be dominated by negative radial velocity caused by the colliding cold pool (e.g. between cold pool A and B in Fig. \ref{fig:cpconstraints}).
\noindent
The presence of neighboring cold pools is also the main motivation for the final two constraints:
\begin{enumerate}[resume]
    \item \textit{The COM of other cold pools cannot be located in the interior of the cold pool in question}.
\end{enumerate}
If the edge of cold pool B in Fig. \ref{fig:cpconstraints} was determined at the far away edge of cold pool C, the COM of cold pool C would be in the interior of cold pool B. 
This is avoided with constraint 3.
\begin{enumerate}[resume]
    \item \textit{The cold pool interior should consist only of positive $v_{r}$:}
    \begin{equation}
        v_{r}(r) > 0 \hspace{0.2cm}\forall\hspace{0.2cm} r \le r^*
    \end{equation}
\end{enumerate}
The purpose of this constraint is equivalent to that of constraint 3 but it is necessary for all slices where the COM of adjacent cold pools are not directly located in the slice.

\noindent
\subsection{Performance evaluation}\label{sec:perf_eval}
\noindent
In order to assess the ability of the tracking to correctly identify and track cold pools, the average near-surface radial structure of the simulated cold pools is examined and compared to results obtained in earlier studies and specifically to results from DH17.
To proceed, composite statistics are computed, where different thermodynamic and dynamical quantities are averaged in both time and space for all identified cold pools:
\begin{enumerate}
    \item For each cold pool, the data at $z = 50$ m ($z = 100$ m for vertical velocity) is interpolated onto a cylindrical grid of ($r, \phi$) with the COM positioned at $r = 0$,
    \item The angular average of the interpolated data across $\phi$ at each r is computed for every time step during the cold pool lifetime,
    \item The resulting radial profiles are averaged at each lifetime for all cold pools.
\end{enumerate}
The main focus of our tracking algorithm is to identify the large, forced updrafts at the cold pool edges. 
Therefore, the average radius of all cold pools at each time during the cold pool lifetime is computed and used as an identifier for the average location of the edge points in the composites.
The average radius is computed as follows:

\begin{equation}
    \langle r_{\tau} \rangle \equiv \sum r_{\tau}/N_{\tau}\;,
    \label{eq:r}
\end{equation}

\noindent
where $r_{\tau}$ is the radius of a single cold pool at a specific lifetime $\tau$, ($r = \sum_{k} d_{k}/n$, where $n$ is the number of edge points identified for the cold pool in question and $d_{k}$ the distance from that cold pools COM to the $k$'th edge point) and $N_{\tau}$ the total number of cold pools at lifetime $\tau$.

\noindent
The average radius is computed both using the tracking algorithm developed in this study and an approximate version of the algorithm developed by DH17. 
Recall that DH17 identified cold pool edges as closed boundaries of the zero contour of $\partial^{2}\theta_{\rho}/\partial r^{2}$. Their edge points are determined by:
\begin{enumerate}
    \item for each cold pool the $\theta_{\rho}$ field at $z = 50$ $m$ is interpolated onto a cylindrical grid of ($r, \phi$), again with the COM defining the origin of $r$,
    \item $\partial^{2}\theta_{\rho}/\partial r^{2}$ is computed to the same accuracy as $\partial v_{r}/\partial r$,
    \item the azimuthal average across all $\phi$ at each $r$ is computed resulting in an average radial profile of $\partial^{2}\theta_{\rho}/\partial r^{2}$,
    \item using polynomial regression a smooth curve is fitted to the averaged radial profile,
    \item the edge is identified by determining the location of zero crossing from positive to negative values, i.e., a local maximum of $\partial\theta_{\rho}/\partial r$,  closest to the center of the smoothed curve ({\it see} Fig. 8 of DH17 for clarity).
\end{enumerate}

\section{Results}\label{sec:results}
\noindent
Generally, larger surface temperature forcing results in a larger number of precipitation events and subsequent cold pools per unit area --- with areal and rain duration cold pool density varying by more than a factor of two for the warmer surface temperature simulations, in principle resulting in substantially altered network of gust fronts for all the different simulations (Tab.~\ref{tab:summary}).
We deliberately developed our tracking method only on one simulation, but then applied it to the others, where $\overline{T_s}$ and $t_0$ were varied, in order to test for robustness.
Sec. \ref{sec:algperf} assesses the algorithm performance by examining results from the simulation performed with p2K ($\overline{T_{s}} = 25 ^\circ C$ and $t_{0} = 1$ day) and Sec. \ref{sec:cpcharac} further examines these results and compares them with results for all remaining numerical experiments.

\begin{figure}[htb]
    \centering
       \includegraphics[width=.9\textwidth]{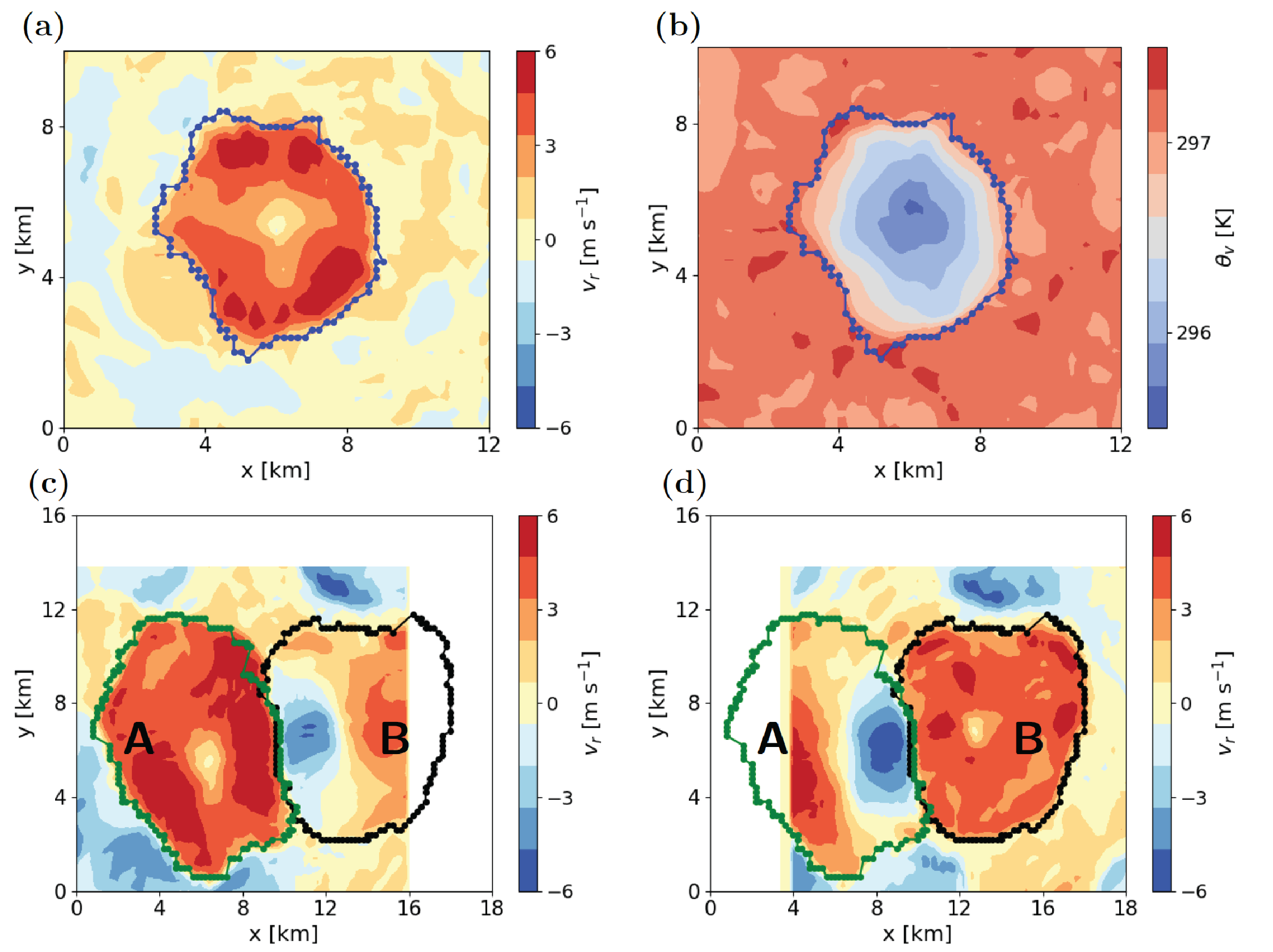}
    \caption{{\bf Detection of cold pool gust fronts.} 
    (a), Example of the identified gust front contour (blue lines and symbols) for a single simulated cold pool in the $| \bm{v}_r|$ field. 
    (b), gust front contour from (a) superimposed onto the $\theta_v$ field. 
    (c), $|\bm{v}_r|$ corresponding to the cold pool labeled A, as well as the detected gust front contour (green lines and symbols). 
    (d), similar to (c), but for the cold pool B. The detected contour of B is shown in black lines and symbols. 
    The white area has no computed values for $\bm{v_{r}}$, because the computation is constrained to the vicinity of the cold pool for increased efficiency. Note the different axis scales in the upper vs. lower panels. Data: simulation p2K (see Sec.~\ref{sec:methods}).}
    \label{fig:cpedges}
\end{figure}


\begin{table}
    \centering
        \begin{tabular}{l | r | r | r | r}
          {\bf Numerical} & {\bf Domain Area} & $\bm{N}_{cp}$ & {\bf Rain duration} & {\bf CP Number Density}\\
            {\bf Experiment} & [$km\times km$] &  & [$hours$] & [$km^{-2}hours^{-1}$]\\
          \hline
          CTR & $205\times 205$ & $444$  & $6.1$ & .0017\\
          p2K & $205\times 205$ & $1389$ & $8.4$ & .0039\\
          p4K & $192\times 192$ & $1470$ & $9.8$ & .0041\\
          LD  & $102\times 102$ & $495$  & $15.2$ & .0031
        \end{tabular}
        \vspace{.5cm}
        \centering
        \caption{{\bf Summary of domain sizes and cold pools detected.} Experiment names are as explained in Sec.~\ref{sec:methods}. ${N}_{cp}$ is the total number of cold pools detected in each simulation. Rain duration is defined as the time during which cold pools can be produced i.e. from the first identified rain event till the last.
        The cold pool number density is the number of cold pools detected divided by the respective domain area and the rain duration.
        }
        \label{tab:summary}
\end{table}

\noindent
\subsection{Algorithm Performance} \label{sec:algperf}

\noindent
Visually, cold pools can often be distinguished by examining either the $v_{r}$ or virtual potential temperature, $\theta_v$, fields. 
The edges identified by the current method qualitatively constitute a reasonable outer boundary in both cases (Fig. \ref{fig:cpedges}a,b, blue contour line) --- large positive values of $v_{r}$ and low $\theta_{v}$ are seen within the detected edge while the surroundings are characterized by small $v_{r}$ and larger, more homogeneous, temperatures.

\noindent
The initial identification of the rain object, which provides a proper spatial reference, together with the sharp gradients in $v_{r}$ at interfaces between closely positioned cold pools, allows the algorithm to both clearly identify single cold pools and distinguish distinct cold pools from one-another (Fig. \ref{fig:cpedges}c,d). 
Together with the fact that the algorithm determines the exact horizontal coordinates of the gust fronts, our findings lead us to conclude that the algorithm is suitable for studies of cold pool collisions.

\begin{figure}[htb]
    \centering
       \includegraphics[width=.6\textwidth]{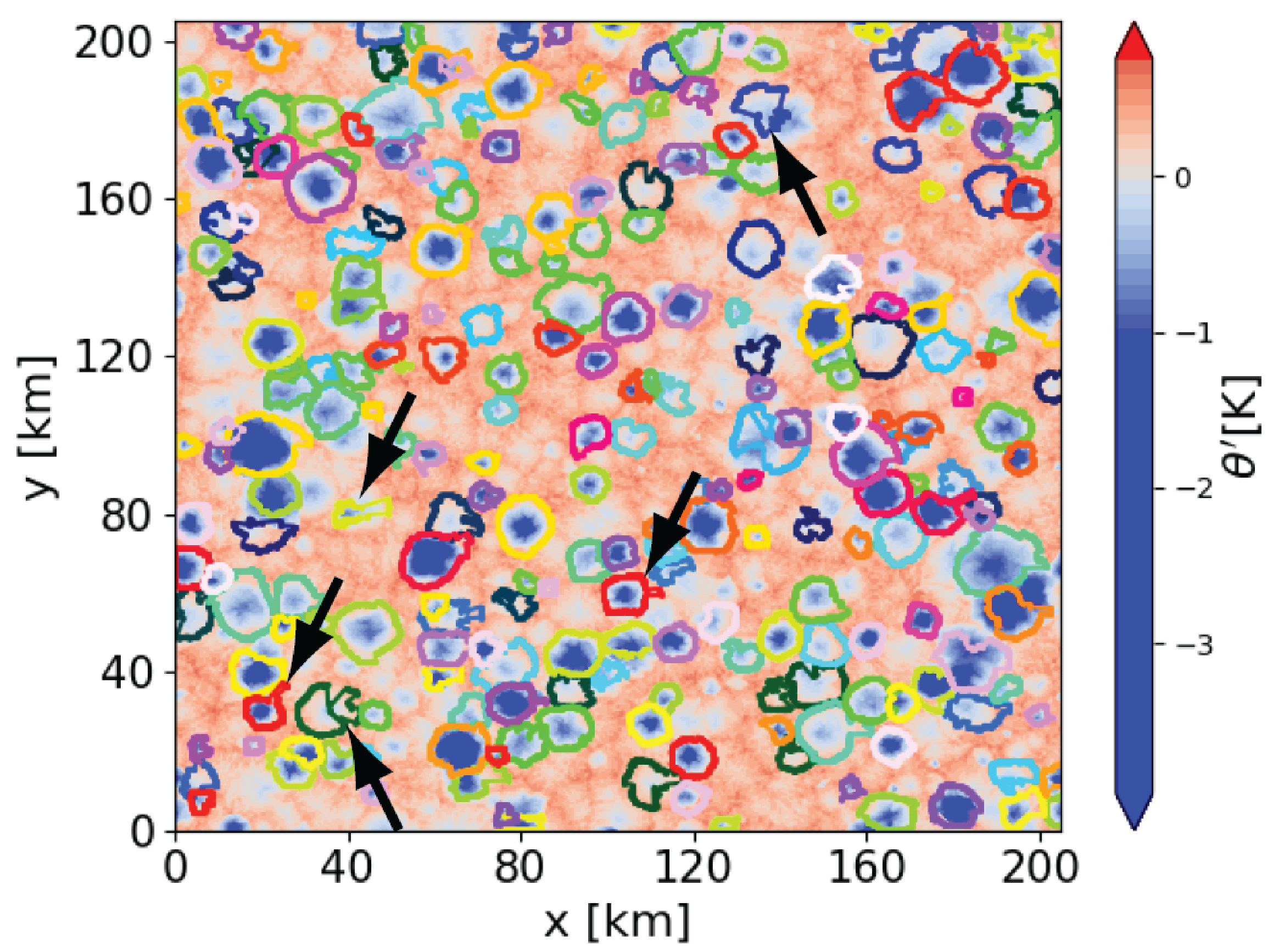}
    \caption{{\bf Example of simulated cold pools with algorithm pitfalls marked.} Color shading ({\it see:} color bar) shows surface potential temperature anomaly. 
    Identified cold pool edges are shown as arbitrarily colored contours. 
    Downwards pointing arrows indicate cold pools with potentially erroneously identified edges. 
    Upwards pointing arrows indicate cold pools where the edges have been identified very near the COM. Data: simulation p2K (see Sec.~\ref{sec:methods}).}
    \label{fig:allcp}
\end{figure}

\noindent
Occasionally, cold pool gust fronts appear to be identified incorrectly.
This occurs most frequently late in the cold pool lifetime where the gradients in the $v_{r}$ field have weakened due to friction and turbulent mixing with the environment. 
Multiple minima in $\partial v_{r}/\partial r$ of comparable sizes can result in identification of edges at intuitively wrong locations ({\it example:} Fig. \ref{fig:allcp}, downwards pointing arrows --- a sudden "jump" of several kilometers is visible in the identified edge contour surrounding the cold pools).
As we will discuss below, the occurrence of detection uncertainty during the late stage of a cold pool's lifetime could be less detrimental to the analysis of dynamical effects --- such as triggering of new convective cells, which is expected to be more likely when the gust front momentum is larger.   

\noindent
Additionally, a large gradient can occasionally exist in the interior of the cold pool due to the COM not being identified directly in the center of the radial expansion or because the cold pool experiences multiple centers, i.e. multiple locations of intense rainfall. 
This can result in the algorithm identifying the cold pool edge very close to the COM ({\it example:} Fig. \ref{fig:allcp}, upwards pointing arrows). 
Future work could explore improvements regarding the identification of the cold pool COM, e.g., by basing it on the wind field generated by the cold pool.

\noindent
If one is interested in the general evolution and structure of cold pools, the errors introduced by the weak gradients are averaged away since the fraction of erroneously identified cold pools appears to be quite small (Fig. \ref{fig:allcp}).

\noindent
\subsection{Cold Pool Characteristics} \label{sec:cpcharac}

\begin{figure}[htb]
    \centering
   \includegraphics[width=.9\textwidth]{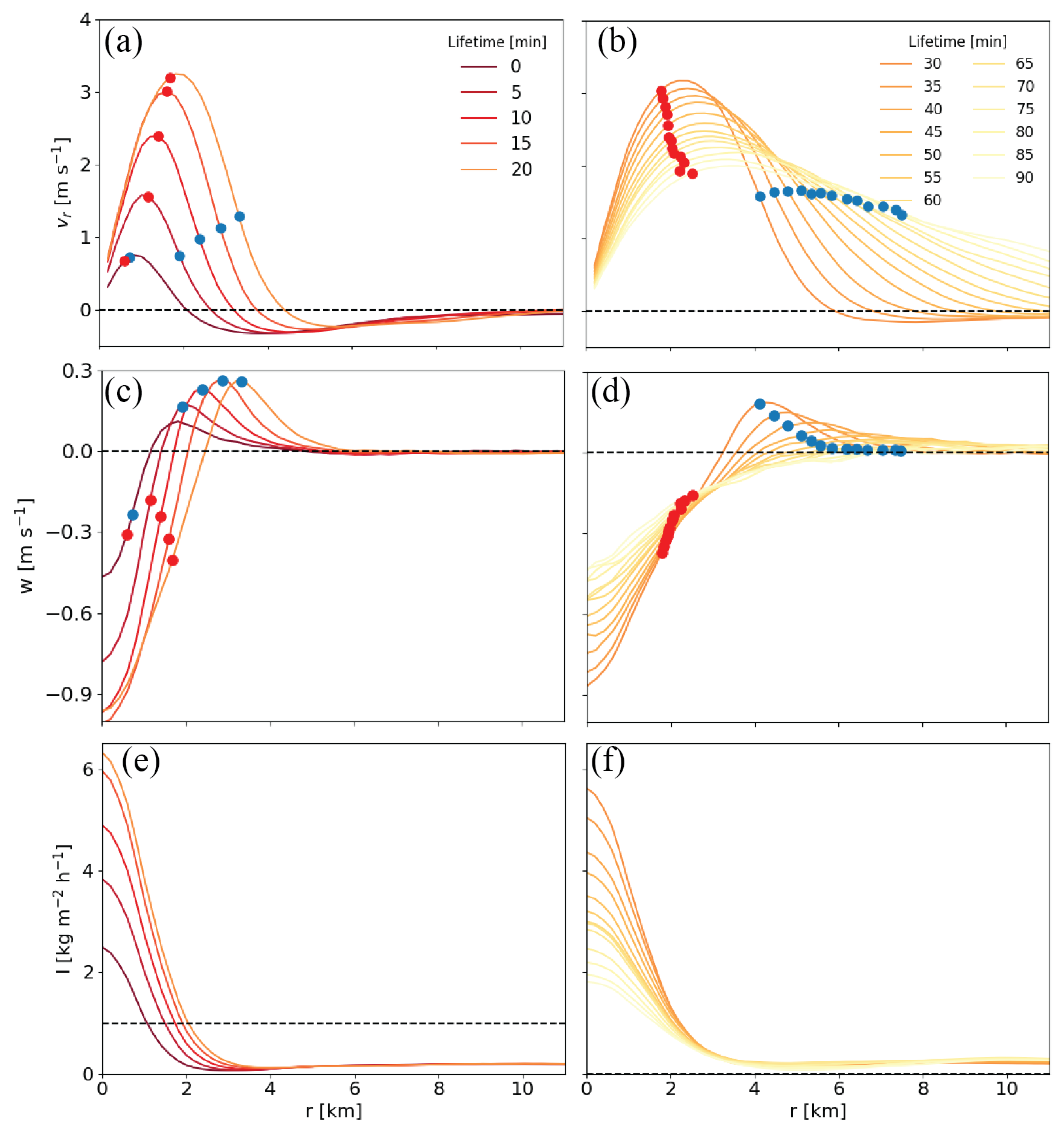}
    \caption{{\bf Temporal and radial evolution of radial and vertical velocity and rain intensity.} 
    {\bf (a)}, surface $v_r$ as function of $r$ at various instances during the cold pool growth phase (see legend).
    {\bf (b)}, similar to (a), but during the cold pool dissipation phase.
    Blue and red circles indicate the averaged radial position of the edges determined by our algorithm and that of DH17, respectively. 
    {\bf (c)/(d)},{\bf (e)/(f)}, Similar to (a)/(b), but for vertical velocity at $z = 100$ m and surface rain intensity, respectively.
    The dashed line in (e) represents the precipitation cutoff.
    Note that, due to possible deformations of cold pools during collisions, the azimuthal averaging used in (a)---(d) is expected to lead to some smearing out of the peaks, somewhat diminishing the amplitude of the curves. Data: simulation p2K (see Sec.~\ref{sec:methods}).
}
    \label{fig:compos1}
\end{figure}

\noindent
We first aim to compare cold pool gust front detection through dynamics (now termed {\it dynamical edge}) to the detection through thermodynamics (termed {\it thermodynamic edge}).
Recall that our algorithm determines cold pool edges through steep gradients in $v_r$ (Fig.~\ref{fig:compos1}a,b). 
As expected by continuity (Eq.~\ref{eq:masscon}), these detected edges correspond to peaks in vertical velocity (blue symbols in Fig. \ref{fig:compos1}c,d).
The thermodynamic edges, in contrast, are located further towards the cold pool COM. The vertical velocities found at these edges and just downwind of them are in fact negative, suggesting subsiding and hence stable, conditions there.

\noindent
Using our condition on dynamics yields detected cold pools that expand during their entire lifetime (Fig. \ref{fig:compos1}a,b),  consistent with the setup of the tracking only recording cold pools while they are {\it dynamically active} ({\it see} Sec. \ref{sec:algo}). 
The dynamical edges are found at or very close to the maximum negative gradient of $v_{r}$ while the thermodynamic edges align well with the peak in $v_{r}$.
The peak in $v_{r}$ should, however, be located upwind from the cold pool edges, while the front is characterized by rapidly decreasing $v_r$. 
Additionally, the density difference between the front and the environment induces a vorticity perturbation acting to increase the surface horizontal winds behind the front and the vertical winds at the front ({\it see:} e.g., Fig. 29 in \citet{wakimoto1982life}).
In the following, we distinguish a growth and a dissipation phase, characterized by times where the intensity of the cold pool parent rain event increases or decreases. 

\noindent
{\bf Dynamics.} We first consider dynamical features:
The growth phase roughly corresponds to times when cold pool expansion accelerates, i.e. the peak $v_{r}$ increases (Fig.~\ref{fig:compos1}a,e), while the opposite is the case during the dissipation phase (Fig.~\ref{fig:compos1}b,f).
These findings are explained by the increasing cooling near the COM during the growth phase, causing increasing gravitational forcing there --- and vice versa for the dissipation phase.
During both phases, the radial velocity at the dynamical edge is of similar magnitude ($v_r\approx 1$ --- $1.5\;m\;s^{-1}$), and remarkably constant during the dissipation phase. 
However, vertical velocities during the growth phase can be a factor of three larger than for comparable radial velocities during the dissipation phase ({\it compare:} Fig.~\ref{fig:compos1}c,d).
Also this feature can be made plausible, when considering that velocities near the COM continue to increase, hence forcing more mass outward. 
Mass conservation (Eq.~\ref{eq:masscon}) must then imply appreciable vertical mass fluxes to make up for the increasing forcing during the growth phase. 
This is an interesting finding, as it might imply that dynamical triggering of new convective events should be expected during the growth rather than the dissipation phase --- hence, early in the cold pool life cycle. 

\begin{figure}[htb]
    \centering
       \includegraphics[width=.9\textwidth]{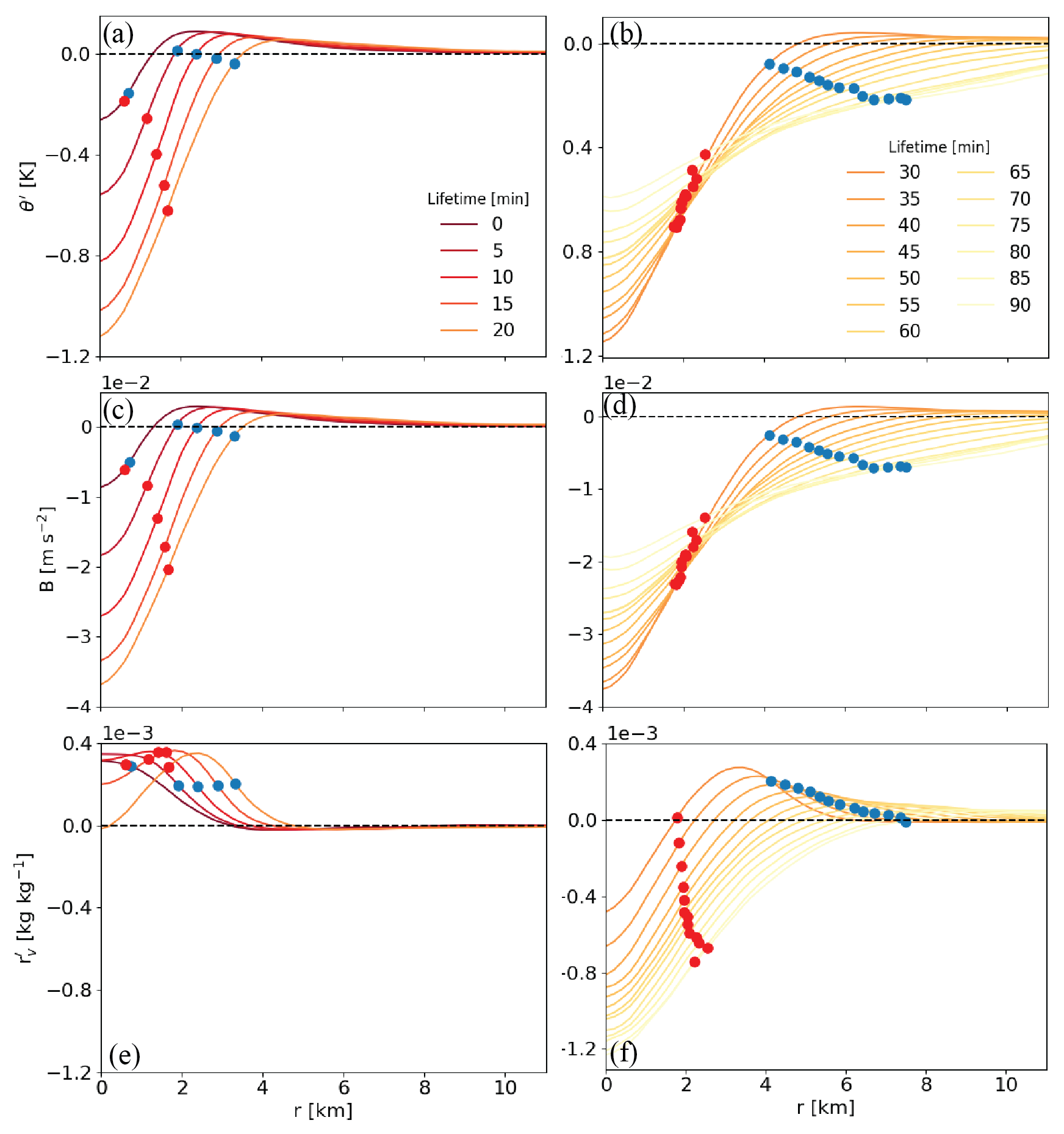}
    \caption{{\bf Temporal and radial evolution of potential temperature, buoyancy and water vapor mixing ratio.} 
    All quantities are shown for the lowest model level, i.e. near the surface. 
    (a),(b), Potential temperature;
    (c)/(d), Buoyancy;
    (e)/(f), Water vapor mixing ratio.
    The presentation is otherwise similar to that in Fig.~\ref{fig:compos1}.}
    \label{fig:compos2}
\end{figure}

\noindent
{\bf Thermodynamics.} Cold pools are typically characterized by negative buoyancy and negative temperature anomalies in their interior (Fig. \ref{fig:compos2}). 
As expected by their definition, the thermodynamic edges are located near the maximum positive gradient in both temperature and buoyancy (Fig. \ref{fig:compos2}a---d) --- 
and generally still negative buoyancy at appreciable distances (1---2 $km$) downwind from the thermodynamic edges.
In contrast, dynamical edges are associated with more modest buoyancy.
In the growth phase (Fig. \ref{fig:compos2}a,c,e) the dynamical edge constitutes an almost perfect demarcation between negative buoyancy, within the detected dynamical edge, and positive buoyancy, surrounding this edge. 
In the dissipation phase (Fig. \ref{fig:compos2}b,d,f), the buoyancy is negative essentially throughout, a feature attributable to the advection of the negative temperature anomaly in the COM of each cold pool, caused by the respective parent rain event.

In both phases, the detected cold pools are surrounded by a band of positive water vapor anomaly, which is advected radially outwards as the cold pool expands, while the center becomes increasingly dry (Fig. \ref{fig:compos2}e,f). 
This is consistent with \citet{tompkins2001organization}, who attributes the drying in the center to the transport of dry air from above cloud base by downdrafts. 
During the dissipation phase, the cold pool signal in all variables gradually fades.
Inspecting Fig.~\ref{fig:compos2}b,d,f, it is worth pointing out that the dynamical edge is still associated with positive moisture anomalies, while the thermodynamic edge occurs at dry locations. 
This may be due to moist sub-cloud air, resulting from rain evaporation, which is quickly advected towards the dynamical edge and makes for a measurable moisture increase there \citep{tompkins2001organization,torri2016rain}. 

\noindent
{\bf Features independent of boundary conditions.}
Our cold pool tracking allows us to compare cold pool characteristics under different boundary conditions (Sec.~\ref{sec:methods}).

\begin{figure}[htb]
    \centering
\includegraphics[width=.6\textwidth]{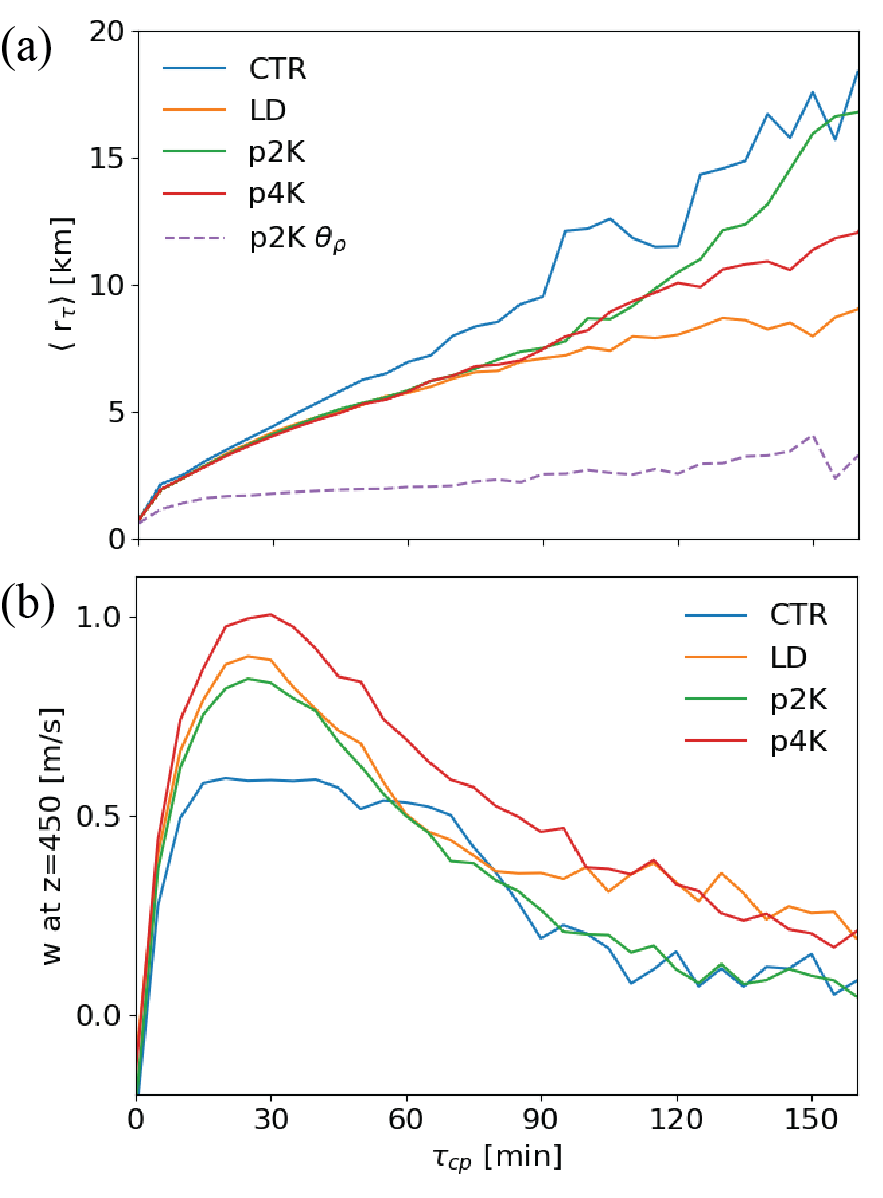}
    \caption{{\bf Temporal evolution of mean cold pool radius and vertical velocity.} 
    Both quantities are measured at $z = 450$ $m$ for all identified edges. 
    Time is measured in terms of cold pool lifetime.
    Note that $\tau_{cp}$ expands further in CTR than in the other numerical experiments, which is explained by the significantly lower cold pool density for CTR ({\it see} Tab. \ref{tab:summary}), allowing cold pools to expand further in CTR before colliding.}
    \label{fig:radevol}
\end{figure}

\noindent
{\bf Cold pool expansion.}
First, we consider the evolution of the mean radius (Fig.~\ref{fig:radevol}a). 
In all experiments, expansion is initially quite rapid ($\approx 10 km\;h^{-1}$), but settles to nearly constant, more modest, speed for all simulations after few minutes. 
As most of the cold pool lifetime occurs during the dissipation phase, the finding of near-constant expansion speed is in line with near-constant $v_r$ during the dissipation phase (Fig.~\ref{fig:compos1}b).
The behavior in all four simulations is rather similar, with no systematic deviation between the curves. 
In terms of updraft velocities (Fig.~\ref{fig:radevol}b), all simulations show a clear peak near the time of peak precipitation intensity ($\approx 30\;min$).

An emergent finding from our analysis hence is: 
for all boundary conditions alike, dynamical triggering effects may be expected to be most pronounced approximately $30$ $min$ after cold pool initiation, at which time cold pools have spread to a $5$ $km$ radius, are neutrally buoyant at the front (Fig.~\ref{fig:compos2}c,d) and have appreciable positive moisture anomalies there (Fig.~\ref{fig:compos2}e,f).  
Using the thermodynamic edge ({\it see} Fig.~\ref{fig:radevol}a, dashed line), many of these triggering effects may not be detectable, as the thermodynamic edge becomes all but stagnant after approximately $10$ $min$.
In our algorithm we were able to detect cold pools up to ages of $90$ $min$, we however caution that the signal-to-noise ratio of the dynamical quantities becomes weaker for even older cold pools.

\begin{figure}[htb]
    \centering
\includegraphics[width=.9\textwidth]{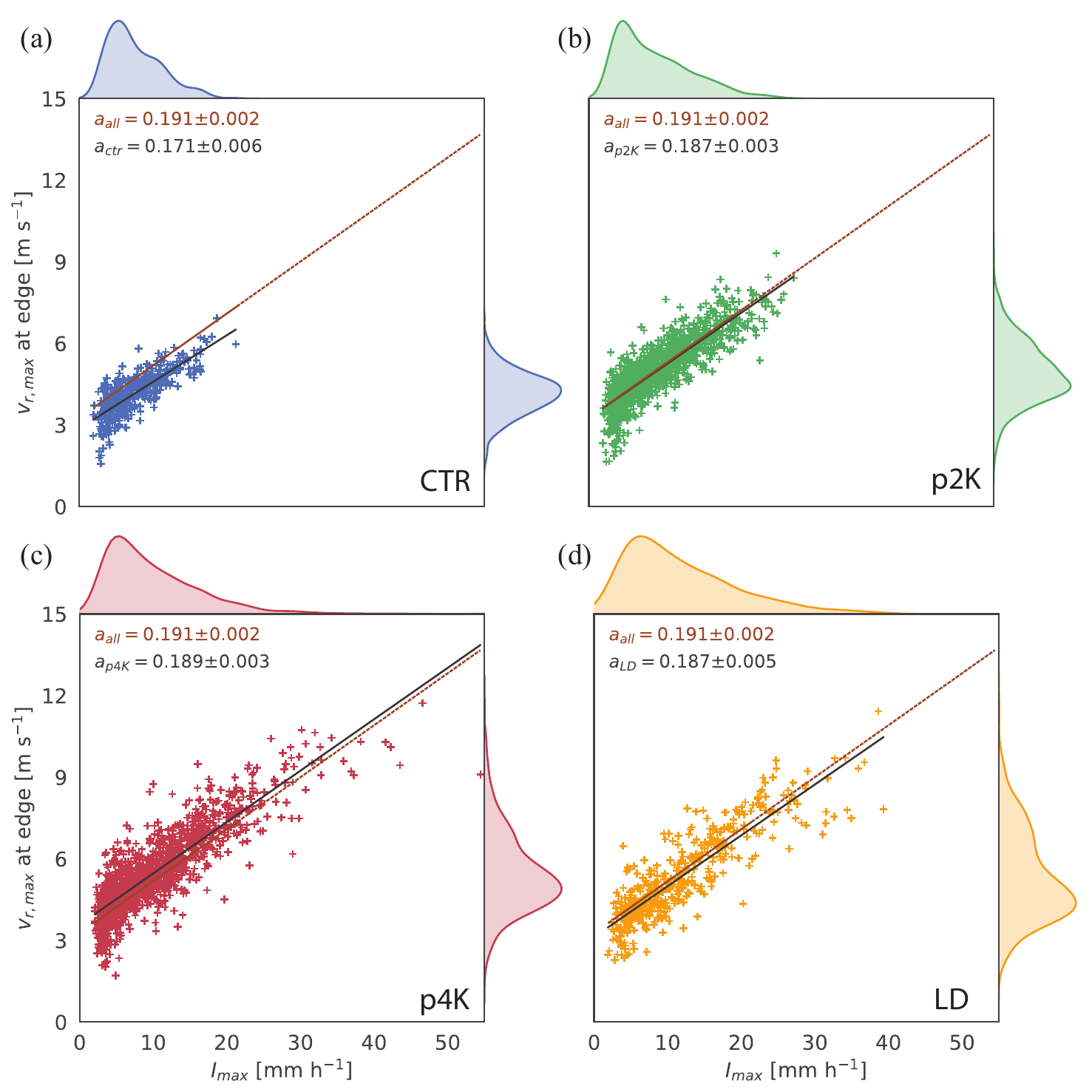}
    \caption{{\bf Relation between maximum precipitation intensity and radial velocity.}} Panels ({\bf a})---({\bf d}) show scatter plots for the four simulations, as labeled, where each symbol represents a single cold pool. 
    Solid black and red lines represent linear fits to the respective individual data and a fit to all data combined, respectively.
    Coefficients in the top left corner of each panel denote the overall slope and slope for the individual simulation, respectively, in units of $m\;s^{-1}h\;mm^{-1}$.
    Shaded curves along the horizontal and vertical axes of each panel indicate the normalized histograms of $I_{max}$ and $v_{r,max}$ corresponding to each experiment.
    \label{fig:vr_vs_Imax}
\end{figure}

\noindent
{\bf Relation between rain events and their cold pools.}
For many applications, such as the parameterization of cold pool in GCMs, it may be useful to obtain generic relations between precipitation events and the resulting cold pools. 
Recently, it was reported that large-eddy simulated convective precipitation cells show rather generic statistical relations, such as a proportionality between event precipitation intensity and event effective radius \citep{moseley2019statistical}.
Using the current cold pool tracking, one can now relate the characteristics of the precipitation event to those of the cold pool for the different numerical experiments. 
The maximal spreading velocity at the dynamical edge, $v_{r,max}$, is a good measure of a cold pool's kinetic energy.
For all simulations, $v_{r,max}$ shows a similar, and nearly linear dependence on maximal precipitation intensity $I_{max}$ for the parent precipitation event (Fig.~\ref{fig:vr_vs_Imax}).
The analysis shows, that larger surface temperature forcing (p2K or p4K) or forcing that is applied over a longer duration (LD) lead to heavier precipitation events, which in turn generate more rapid cold pool expansion --- however, the relation between the two quantities remains all but unchanged within the four different sensitivity experiments. 
With the proportionality of event effective radius $r_{eff}\sim I_{max}$ \citep{moseley2019statistical}, a square root dependence between event area and $v_{r,max}$ is further implied.
We find that this holds reasonably well, when using the maximum event area $A_{max}$ of the parent rain event, that is, $v_{r;max}\sim A_{max}^{1/2}$.  

\section{Summary and conclusion}\label{sec:conclusion}
\noindent
We have presented a cold pool tracking algorithm, which detects the {\it dynamical edge}, one characterized by the convergence lines surrounding each cold pool, after the cold pool has been generated by a precipitation event. 
The dynamical edge is distinct from the edges identified in previous cold pool tracking methods \citep{drager2017characterizing}, where a temperature-based edge has been employed.  
The motivation for using such dynamical edges is the finding that updrafts often result in regions of strong near-surface convergence, and dynamical triggering of new precipitation events is expected to be facilitated at these locations. 

Due to entrainment and turbulent mixing of environmental air, this dynamical cold pool edge can exhibit thermodynamic properties quite similar to those of the environmental air in the surroundings, making the gust front all but undetectable through thermodynamic approaches. 
For thermodynamic edges, which we track for comparison, our findings conversely indicate pronounced stability due to downdrafts and relatively cool boundary layer conditions. 
Thermodynamic edges may hence be useful in describing areas where convection is suppressed. 
In future studies on cold pool-precipitation dynamics, it could hence be beneficial to combine both approaches to characterize cold pool regions of enhanced stability and enhanced triggering potential. 

To summarize, our algorithm tracks cold pools throughout their lifetime. 
The method is simple, as it requires only the
tracks of surface precipitation, yielding precipitation-weighted center of mass coordinates for each time step and rain cell, as well as the two-dimensional near-surface horizontal wind field.
For the unit circle surrounding a given precipitation area center of mass, the tracking breaks down the azimuthal range into slices of equal angular range. 
Within each slice a maximum of radial velocity change is identified --- corresponding to radii of maximum convergence. 
Several checks are applied at each timestep, to ensure that noise does not strongly perturb the detection of each cold pool gust front. 

Our tracking successfully detects an ensemble of cold pools generated by running large eddy simulations (LES) for a diurnal cycle case, initialized with soundings from mid-latitude potentially convective summer days (Sec.~\ref{sec:methods}). 
In order to assess the performance of the current dynamic-based tracking algorithm in relation to a thermodynamic-based one, a comparison with the tracking algorithm developed by \citet{drager2017characterizing} is performed. 
The average edge location for all cold pools at each time during their total lifetime is computed for both methods. 
In general, the edges determined using the dynamical edge are located radially further away from the cold pool center than the ones determined using the thermodynamic method. 
As the present method involves checks to ensure that one cold pool edge must not be located within the interior of another cold pool, a general overestimation of the radii detected by the present method can be excluded. 
In conclusion, our results suggest that the dynamics-based method allows for more complete tracking of the area enclosed by each cold pool gust front. 

In practice, our analysis shows that after approximately 30 minutes the thermodynamic edges all but cease to advance further, only reaching a maximum radius of approximately two kilometers, whereas the dynamical edges continue to advance (Fig.~\ref{fig:radevol}a).
This finding has implications when examining the dynamics within the interior of cold pools, at positions between the center and the front (Fig.~\ref{fig:compos1}) as well as the average radial structure of temperature, buoyancy, moisture and vertical velocity (Fig.~\ref{fig:compos2}). 
The dynamical edges collocate with only weakly negative, or even neutral, buoyancy and generally positive moisture anomalies. 
As the gust fronts detected are further located near updraft maxima, they likely form "hotspots" for triggering of new precipitation events. 

We exemplify the scope of our method by drawing a linear relation between event maximum precipitation intensity and the peak expansion speed of the resulting cold pool --- a relation that is nearly unchanged for the different simulations. 
Such relations may be a first step towards mechanistic parameterization of cold pool dynamics in general circulation models. 
A possible challenge might be strong wind shear, which could change the shape and symmetry of cold pools and the location of the rain event COM relative to the gust front. 
In the specific case of squall lines it could be of interest to only consider the cold pool loci with the largest gradient in radial velocity, often facing the direction of squall line propagation. These loci might locally still be approximated by  circle segments, therefore still allowing them to be identified by our algorithm. 
Another challenge not addressed here is the case of merging cold pools where the corresponding rain events do not merge.
These challenges should be addressed in a subsequent paper.

Despite the limitations discussed, our algorithm could in principle be used to track cold pools in observational data. 
Rain tracks can be easily measured by radar but records of wind speed at high spatial resolution are currently not common. 
A useful setup for field studies could be to select an area of at least $10 km \times 10 km$ with frequent occurrence of convective events, homogeneous surface conditions, and a network of narrowly spaced (kilometer scale) wind measurements.

Needless to say, further analysis should now follow. 
It is important to close the precipitation-cold pool feedback loop by detecting the influence of cold pool gust fronts on the generation of new precipitation events.
Attempts have been made at describing self-organization through cold pools in conceptual models \citep{grandpeix2010density,boing2016object,haerter2018intensified,haerter2019circling}.
With more mechanistic information in place, a full, physics based, cellular automaton, mimicking the self-organization of convective cold pools in space and time, could be built. 
Beyond this, the details of convective self-organization, such as clustering and the formation of extreme events, could be deciphered from analysis that builds on the current method.

\acknowledgments
MBF and JOH gratefully acknowledge funding by a grant from the VILLUM Foundation (grant number: 13168) and the European Research Council (ERC) under the European Union's Horizon 2020 research and innovation program (grant number: 771859).
The LES simulation data used in the this study are available from \citet{moseley2016intensification}.
The authors are grateful for computing resources and technical assistance provided by the Danish Center for Climate Computing, a facility built with support of the Danish e-Infrastructure Corporation, Danish Hydrocarbon Research and Technology Centre, VILLUM Foundation, and the Niels Bohr Institute.
The current tracking code is available at \url{https://github.com/mariellebf/cp_tracking}, a user manual is attached as supplementary material.
No new data were produced in this study.


%


\end{document}